\documentclass[final,english]{bullsrsl}[2022/06/15]

\usepackage[latin1]{inputenc}
\usepackage[T1]{fontenc}

\usepackage{natbib} 
\usepackage{graphicx}

\begin{document}
\title{Caught in flagrante delicto: evidence for past mass transfer in massive binaries?}

\author{Gregor}{Rauw}
\affiliation{STAR Institute, Universit\'e de Li\`ege, All\'ee du 6 Ao\^ut, 19c, B\^at B5c, 4000 Li\`ege, Belgium}
\correspondance{g.rauw@uliege.be}
\date{5 June 2024}
\maketitle

\begin{abstract}
Many massive binary systems undergo mass and angular momentum transfer over the course of their evolution. This kind of interaction is expected to deeply affect the properties of the mass donor and mass gainer and to leave various observational signatures. The most common smoking guns of a past mass transfer episode are notably rapid rotation of the mass gainer and altered surface chemical abundances of the stripped mass donor star. Quantitative observational studies of evolved massive binaries are crucial to gain insight into poorly constrained parameters of binary evolution models such as the fraction of mass lost by the mass donor that is actually accreted by the mass gainer. Yet, drawing conclusions about a past mass transfer episode requires a detailed analysis of all aspects of a binary system which sometimes leads to unexpected results. In this contribution, we review the existing observational evidence for past mass exchange events in massive main-sequence and post main-sequence binaries.
\end{abstract}


\begin{altabstract}
  \textbf{Prises en flagrant d\'elit: signatures d'\'episodes pass\'es d'\'echanges de mati\`ere dans des binaires massives?} Beaucoup de syst\`emes binaires massifs traversent des \'episodes de transfert de masse et de moment angulaire au cours de leur \'evolution. Ce type d'interaction affecte les propri\'et\'es des deux \'etoiles (donneur de masse et accr\'eteur) et devrait impr\'egner plusieurs signatures observationnelles. Parmi les indices les plus probants d'un \'episode pass\'e d'\'echange de mati\`ere, il y a la rotation rapide de l'accr\'eteur et la composition chimique alt\'er\'ee du donneur de masse. Des \'etudes observationnelles quantitatives de binaires massives \'evolu\'ees sont cruciales pour contraindre certains param\`etres cl\'es des mod\`eles d'\'evolution, tels que la fraction de la mati\`ere perdue par le donneur qui est effectivement accr\'et\'ee par le compagnon. Toutefois, ce type d'\'etude requiert une analyse en profondeur de tous les aspects d'un syst\`eme binaire, ce qui conduit parfois \`a des r\'esultats inattendus. Dans cette contribution, nous passons en revue les indices observationnels d'\'episodes pass\'es d'\'echanges de masse dans des binaires massives de la s\'equence principale ou ayant quitt\'e celle-ci r\'ecemment.    
\end{altabstract}

\altkeywords{\'etoiles massives, binaires s\'err\'ees, \'etoiles: abondances, \'etoiles: \'evolution, \'etoiles: rotation}

\section{What theoretical models predict \label{theory}}
Contrary to single star evolution, which is mostly ruled by the initial mass of the star, its rotation rate, wind mass-loss rate, overshooting prescription and metallicity \citep[e.g.,][]{Eks11,Mar13}, the evolution of a massive binary system depends on a number of additional parameters, such as the orbital period, the primary/secondary mass ratio, and the orbital eccentricity. As a result, a wide variety of binary evolutionary channels can be distinguished \citep[e.g.,][]{Pod92,Chen24}. These scenarios can be classified into three broad categories depending on the evolutionary phase of the initially more massive star at the time of the mass-transfer \citep{Kip67,Vanb98,MarBod23}. If the mass donor is in the hydrogen core-burning phase, one speaks about Case A mass transfer. Case B concerns situations where the mass exchange occurs after the donor has completed its core hydrogen burning phase but prior to the ignition of core helium burning. Finally, Case C refers to situations where the primary has evolved beyond helium core-burning. About a quarter of the massive binaries have sufficiently short orbital periods to ensure that mass transfer occurs while the stars are still in the core hydrogen burning phase \citep{deM14,Sen22}. For a 25\,M$_{\odot}$ primary, Case A mass transfer occurs in systems with initial $P_{\rm orb}$ below about 10\,days \citep{Lan20,Sen22}. In wider systems, mass transfer occurs as Case B.

To guide the theoretical binary evolution models, it is crucial to collect as many observational diagnostics as possible for each of the evolutionary stages massive binaries go through. In some cases, we can directly observe the mass exchange taking place between the stars. In a Case A situation, the orbit shrinks as the more massive star expands and initiates the Roche lobe overflow (RLOF) process. This leads to a fast mass transfer episode occuring on the thermal timescale of the mass donor \citep[e.g.,][]{Sen22,MarBod23}. Once the mass ratio is inverted, the orbit widens. The second Case A mass exchange episode is then expected to be much slower as it occurs on the much longer nuclear timescale. Given the difference in timescale between the fast and slow Case A, the majority of the observable semi-detached systems, where mass exchange can be seen directly, are expected to be in the slow Case A phase. These systems, where the mass donor is now the less massive but more evolved star, are called massive Algol binaries. Some of these systems, those with very short orbital periods, enter a contact configuration during the slow Case A mass transfer and eventually experience mass loss through the L$_2$ point leading to significant losses of angular momentum and finally to the merger of the two stars.

In very massive binaries, the fast Case A mass transfer can end before the mass ratio has inverted \citep{Sen23}. This occurs when the mass donor is stripped sufficiently for He-rich material to appear at the surface. Subsequent mass loss then leads to a shrinking of the donor. Moreover, wind mass loss increases and becomes so strong that the orbit widens already before the mass ratio is inverted \citep{Sen23}. During the ensuing nuclear timescale evolution, many mass donors remain more massive than the accretors. Such systems appear as so-called reverse Algol systems. Actually, the slow (nuclear timescale) mass transfer may be interrupted or absent altogether. This scenario can lead to the formation of Wolf-Rayet (WR) binaries containing H-rich WN stars as the mass donors.

Observational diagnostics are however not restricted to the massive Algol or reverse Algol systems. The consequences of past RLOF episodes can also be diagnosed in other binaries. Moreover, in many of the binary evolutionary scenarios, a fraction of the binary systems either merge or break-up. Therefore, the signatures of past binary interactions also affect the properties of currently single stars. In general terms, binary evolution is expected to result in various observational signatures that we briefly discuss hereafter:
\begin{itemize}
\item[$\bullet$] spin-up of the mass gainer,
\item[$\bullet$] altered chemical composition at the surface of the mass donor,
\item[$\bullet$] overluminosities of the mass donor compared to main-sequence stars of same mass,
\item[$\bullet$] and possibly magnetic fields. 
\end{itemize}

Roche lobe overflow not only leads to the exchange of mass between the stars, but also to the transfer of angular momentum from the mass donor to the mass gainer. How this affects the accretion process depends on the efficiency of tidal interactions. Indeed, when the gainer is spun up to critical rotation it stops accreting \citep{Pac81}, leading to a low mass transfer efficiency. Tides tend to synchronize the rotation and the orbital motion and thus to spin down the mass gainer, allowing for a more efficient mass transfer \citep[e.g.,][]{Lan20,Sen22}. Case A close binary systems evolve quickly into tidal locking and should thus have a high mass transfer efficiency. Conversely, tidal effects are negligible in wider systems. Case B systems, where the companion quickly reaches critical rotation, should thus have a low mass transfer efficiency. The spun-up mass gainer might then become an Oe/Be star \citep[or a 'normal' rapidly rotating OB star;][]{VanBever} paired with a hot He-burning subdwarf companion, a white dwarf, or a neutron star \citep[e.g.,][and references therein]{Shao21}.\\
Another way to produce a rapidly rotating massive star could be through a binary merger: \citet{deM13} argue that binary interactions account for 20\% of all massive main-sequence stars having projected rotational velocities ($v\,\sin{i}$) exceeding 200\,km\,s$^{-1}$, either as a result of mass transfer or binary mergers. However, \citet{Sch19} rather predict that merger products, once they return to thermal equilibrium and settle on the main-sequence, should be rotating slowly. 

The stripping of the mass donor reveals material at the stellar surface that was previously inside the convective core of this star. Therefore, one expects the mass donors to display enhanced nitrogen and helium abundances, along with depleted carbon and oxygen at their surface \citep[e.g.,][]{Sen22}. Binaries that survive the entire Case A mass transfer phase are expected to reach the CNO equilibrium abundances. In single star evolution, rotational mixing plays a key role in chemical enrichment at the stellar surface. In contrast with this, in binary evolution, where the surface enrichment is due to envelope stripping, one does not expect a correlation between surface nitrogen enrichment and rotation rate of the donors.\\
The chemical enrichment of the mass gainer depends on the mass accretion efficiency. For instance, in Case B binaries, the mass gainer essentially maintains its original surface composition because it accretes only a small fraction of the material lost by the donor. Furthermore, regardless of the accretion efficiency, the accreted chemically enriched material is diluted by thermohaline mixing with unprocessed material of the accretor. As a result, any chemical enrichment of the accretor is generally expected to remain well below that of the mass donor \citep[e.g.,][]{DeL94,Vanb94,Lan20,Sen22}. Nonetheless, the overall properties (chemical composition and rotation rate) of accretors are clearly distinct from those of single rotating stars \citep[e.g.,][]{Ren21}.

The mass donors of massive Algol systems are expected to be overluminous compared to single stars of same mass and effective temperature \citep{Sen22}. This is a consequence of the modified chemical composition of the stellar surface following the stripping of the hydrogen-rich envelope. Indeed, the stellar luminosity scales as $L \sim M^{\alpha}\,\mu^{\beta}$ with $\mu$ the mean molecular weight and $\beta$ in the range 5 - 2.5 for stars between 10 and 30\,M$_{\odot}$ \citep{Sen22}. The value of $\mu$ increases by about a factor 2.2 during the hydrogen burning phase, thus leading to the overluminosity once the enriched material appears at the surface.\\
Depending on their mass, stripped envelope mass donors that have previously undergone Case B mass transfer should exhibit spectral characteristics ranging from hot subdwarfs to Wolf-Rayet (WR) stars \citep{Vanb91,Goet18}. These objects have their hot helium core exposed with only a thin residual layer of hydrogen on top. \citet{Goet18} found that these stripped stars display bolometric luminosities similar to those of their progenitors despite having lost about two third of the initial mass. This is because of their extreme temperatures that compensate for their small radii. These temperatures shift the peaks of the spectral energy distributions of the stripped stars into the extreme UV domain. 
  
The magnetic fields that are detected in about 7\% of the massive stars might be another consequence of binary interactions. In merger events involving two main-sequence stars, fresh nuclear fuel is mixed into the core of the merger product. This leads to a rejuvenation of the order of the nuclear timescale \citep{Sch16}. The products of such merger events should thus appear younger than other stars that formed simultaneously with their progenitors. Moreover, such events could lead to the formation of strong long-lived magnetic fields \citep{Sch16,Sch19}. The magnetic field would be triggered by the shear of the massive accretion stream on the surface of the accretor, and would be further amplified by vortices occuring when the two stellar cores merge \citep{Sch19}.\\
From measurements of projected rotational velocities of O-star components in Galactic and Large Magellanic Cloud (LMC) WR + O binaries \citep{Sha17,Sha20}, \citet{Vanb18} noted that the O-star components of those systems are rejuvenated and rotate faster than synchronously, although significantly slower than their critical velocities. This situation was interpreted as evidence for spin-up by accretion combined with a mechanism to slow down the rotation. Indeed, the tidal interactions alone are not sufficient to slow down the mass gainers from near critical rotation to their current spin rates. \citet{Vanb18} accordingly suggest that accretion leads to differential rotation in the mass gainer, thereby triggering a Spruit-Tayler dynamo mechanism that generates a strong magnetic field. The magnetic field then forces the stellar wind of the accretor and the inflowing material into co-rotation with the star out to the Alfv\`en radius. This leads to a substantial loss of angular momentum and thus a spin-down of the star \citep{udD09}.

\section{The broad observational picture \label{population}}
Systematic observing campaigns of massive stars have shown that a majority of them are located in binary or higher multiplicity systems \citep[][and references therein]{Sana12}. Binary interactions should thus impact the overall properties of many massive stars. In this section, we discuss the observational signatures of RLOF interactions on populations of massive stars. 
\subsection{Fast rotators}
The observed distribution of $v\,\sin{i}$ values of apparently single OB-type stars displays a bimodal shape consisting of a main peak at around 80 -- 100\,km\,s$^{-1}$ flanked by a tail of rapid rotators ($v\,\sin{i} \geq 200$\,km\,s$^{-1}$) extending up to $\sim 400$\,km\,s$^{-1}$ \citep[e.g.,][and references therein]{Ram13,Duf13,Hol22}. The tail of rapid rotators could consist mostly of post-binary interaction products \citep{deM13}. These would be either merger products or rejuvenated mass gainers possibly accompanied by a stripped star or a compact object.

\citet{Bri23} analysed a sample of 54 fast rotating O-type stars comparing it to a sample of 415 Galactic O-type stars spanning the full range of $v\,\sin{i}$ values. They concluded that there is a deficit of double-lined spectroscopic binaries (SB2) among fast rotating O-type stars (8 - 12\% versus $\sim 33$\% for slow rotators). This is a likely consequence of the presence of binary evolution products among the population of rapid rotators. Indeed, many of the rapid rotators in binary systems are likely paired with a stripped companion or a compact object neither of which produces a strong spectral signature. Conversely, the higher fraction of SB2 systems among slow rotators indicates that these systems have not yet undergone binary interaction. Moreover, \citet{Bri23} found that the fraction of runaway stars\footnote{To avoid biases from uncertain radial velocities (RVs), the runaway status was assessed on the tangential velocity only.} is significantly higher among the fast rotators (47\% versus 20 - 30\%). The fraction of runaways is even higher among those fast rotating O-stars that are classified as presumably single stars. In this subsample, 64\% (23 out of 36 objects) were found to be runaways. This result is in line with the interpretation that these rapid rotators were ejected from their birthplace as a consequence of the supernova explosion of a former mass donor in a binary system.

The rapidly rotating O9.5 star HD\,93521, which is located about 1\,kpc above the Galactic plane, could be an example of the outcome of binary evolution. \citet{Gie22} showed that it would require about 39\,Myr for this star to reach its current location starting from the Galactic plane. This is much longer than the estimated age of the star of about 5\,Myr. \citet{Gie22} therefore suggest that HD\,93521 formed via the merger of two stars of nearly equal mass.

Classical Oe/Be stars are rapidly rotating OB stars surrounded by decretion disks. One possibility to explain their near-critical rotation is that the Be stars were spun up by mass and angular momentum accretion during a past RLOF episode in an interacting binary. A significant fraction of the mass donors are expected to be in the He-core burning stage and should appear as slowly rotating stripped hot subdwarf sdO/sdB stars \citep{Goet18}. Such objects are difficult to detect against the strong optical emission of the Be star, and many of them might even escape detection in the UV \citep{Yun24}. Nevertheless, about 20 such sdO companions have been found via cross-correlation with far-UV spectra \citep[][and references therein]{Wan21,Wan23}. These results suggest that many Be stars could have such hot subdwarf companions, thereby supporting the binary scenario.

Since most of the stripped companions of Be stars predicted by theory are expected to be hot subdwarfs they should be too faint in the optical domain to be distinguished against the strong emission of the Be star. The situation could be different for bloated stripped companions, which represent an earlier evolution stage where the mass transfer is possibly still ongoing.
\citet{ElB22} performed a search for such Be + bloated stripped companion systems among the spectra of the APOGEE survey. They searched for Doppler shifts of narrow photospheric absorptions associated with the bloated stripped star. Out of a sample of 297 stars classified as Be stars, they identified one such binary system: HD\,15124 consisting of a Be primary star ($T_{\rm eff} = 16000$\,K) with a mass of about 5.3\,M$_{\odot}$ and a low-mass (0.92\,M$_{\odot}$) cool ($T_{\rm eff} = 6900$\,K) subgiant companion revolving on a 5.47\,days circular orbit\footnote{\citet{ElB22} did not measure the RVs of the Be star. Hence they derived an SB1 orbital solution, and the mass of the primary was inferred from the spectroscopic $\log{g}$ value.}. The light curve of this system revealed a combination of ellipsoidal variations, due to the low-mass companion filling up its Roche lobe, and reflection of the B star light off the surface of the low-mass subgiant. \citet{ElB22} showed that the emission lines in the spectrum arise from an accreting disk fed by the low-mass star which fills-up its Roche lobe, and that the mass gainer currently rotates at only 60\% of its critical velocity. As such, the primary in HD\,15124 is not yet a genuine classical Be star.
\citet{ElB22} further used stellar atmosphere models to constrain the chemical abundances. The He abundance was derived from the spectrum of the primary star and was found to be 4 times higher than solar. The CNO abundances were determined from the spectrum of the mass donor. Both C and O were found to be heavily depleted whilst N was found to be overabundant by a factor 6.

Conversely, fast rotation can also be primordial! Indeed, a group of rapidly rotating OB stars paired with a low-mass (typically 1\,M$_{\odot}$) companion was found to most likely consist of stars that have not yet experienced any mass-exchange \citep[][and references therein]{Naz23}. Three such systems were studied in detail by \citet{Naz23}: HD\,25631 (B3\,V, $P_{\rm orb} = 5.24$\,days), HD\,191495 (B0\,V, $P_{\rm orb} = 3.64$\,days), and HD\,46485 (O7\,V, $P_{\rm orb} = 6.94$\,days). These systems display photometric light curves with a strong reflection effect and two narrow eclipses. Unlike HD\,15124, they do not show strong ellipsoidal variations though, indicating that none of the binary components has a large Roche lobe filling factor. In contrast to systems consisting of Be stars paired with hot subdwarfs, the OB primary star is the hotter component in these systems. The mass ratios are rather extreme \citep[$M_s/M_p$ in the range 0.035 to 0.15 for the three systems studied by][]{Naz23}, and the secondaries are likely non-degenerate, low-mass pre-main-sequence stars. To further understand the status of these systems, \citet{Bri24} performed evolutionary calculations testing different assumptions regarding the origin of the primary star's fast rotation. Assuming a primordial origin, \citet{Bri24} showed that the loss of angular momentum via wind mass-loss becomes only significant towards the end of the main-sequence phase. Likewise, tidal synchronisation also becomes effective only at the end of the main-sequence phase. This implies that stars with initially high rotational velocities maintain this property over most of their main-sequence lifetime. Assuming instead a past binary interaction, \citet{Bri24} showed that during a conservative mass transfer, the orbit widens and the orbital periods of post-RLOF systems are thus much longer than those observed. If the mass transfer is assumed non-conservative, the orbit shrinks, but an extreme mass-ratio configuration (as observed) is not expected in such cases as these systems would merge. Hierarchical triple systems could in principle lead to a fast rotator (resulting from the merger of the inner binary) coupled to a lower mass companion (the tertiary component of the initial triple system). However, the properties of the systems studied by \citet{Naz23} are such that these triple systems would likely be dynamically unstable. \citet{Bri24} thus conclude that the rapid rotation of these stars is primordial.

\begin{figure}
\centering
\resizebox{12cm}{!}{\includegraphics{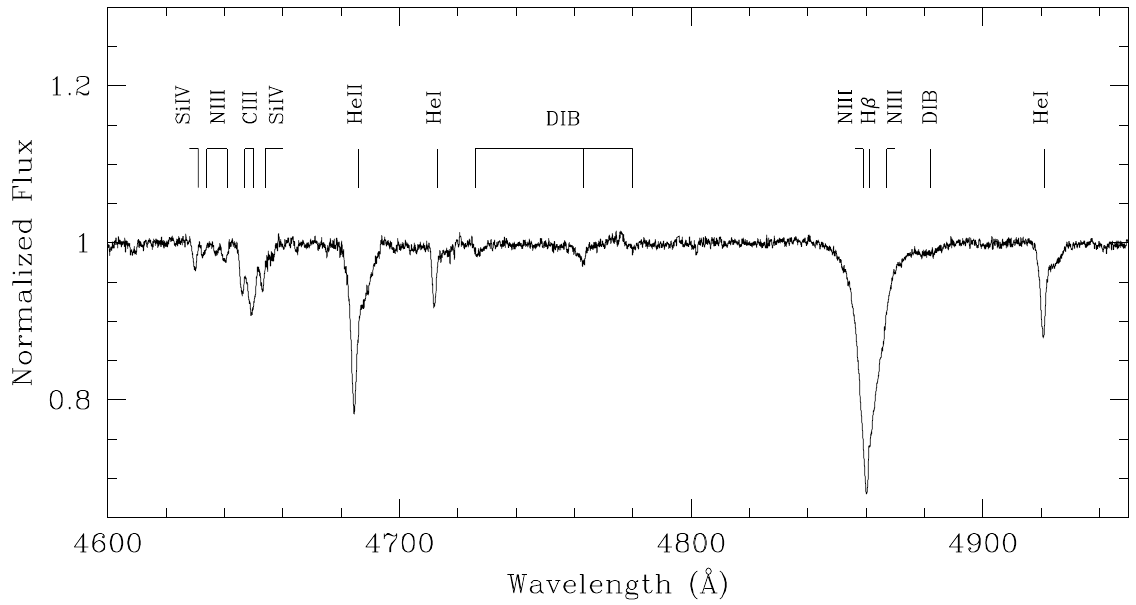}}
\bigskip
\begin{minipage}{12cm}
 \caption{Blue spectrum of HD\,93343 obtained with the FEROS spectrograph at the 2.2\,m telescope of the ESO La Silla observatory in Chile on 25 May 2003. The blend between the narrow absorptions of the primary and the much broader absorptions of the secondary is best seen in the He\,I $\lambda\lambda$\,4713, 4922, He\,{\sc ii} $\lambda$\,4686 and H$\beta$ spectral lines.\label{HD93343}}
\end{minipage} 
\end{figure}

Another striking case of asynchronous rotation that is not due to a mass transfer is provided by the O7.5\,Vz + O7.5\,Vz binary HD\,93343 \citep[$P_{\rm orb} = 50.4$\,days, $e = 0.398$,][]{Put18}. Both stars of this system have essentially identical masses ($M_s/M_p = 0.97 \pm 0.01$, $M_p\,\sin^3{i} = 22.8$\,M$_{\odot}$), but differ strongly by their projected rotational velocities \citep{Rauw09,Put18}: $v_p\,\sin{i} = 65$\,km\,s$^{-1}$ and $v_s\,\sin{i} = 325$\,km\,s$^{-1}$ for the narrow line and broad line components, respectively (see Fig.\,\ref{HD93343}). Whilst such a difference could arise from a past RLOF episode (see the case of Plaskett's Star in Sect.\,\ref{Plaskett} hereafter), the stars in HD\,93343 are very young unevolved objects \citep[as indicated by the `z' tag in their spectral classification,][]{Put18}, and mass exchange has not yet played any role. Hence, the difference in rotation must reflect either a misalignment of the rotation axes or stems from the formation mechanism of this young eccentric binary system. 

\subsection{Chemical enrichment}
Enhanced N and He as well as depleted C and O abundances are expected at the surface of massive stars either as the result of rotational mixing or following the removal of the outer layers either by the action of a stellar wind or by binary interactions. Late O-type stars with unusually strong nitrogen lines in their spectra are classified as ON stars. \citet{Mar15} studied the properties of 12 ON stars which are not members of SB2 binaries. All of these stars are He-rich and display N/C ratios between 0.5 and 1.0 dex higher than in normal O-stars. Though ON stars rotate faster on average than normal O-stars, evolution of single rotating stars cannot account for this enrichment as it is observed even for stars which are still on the main-sequence \citep{Mar15}. Moreover, not all fast rotators display ON spectral characteristics. Some of the stars in the sample are likely members of SB1 binaries, suggesting that the ON stars would be the former mass gainers of these binaries. This is remarkable since in two post-RLOF SB2 systems that host ON stars (LZ\,Cep and HD\,149404, see Sect.\,\ref{How}) the ON component is the mass donor instead.

\citet{Caz17a,Caz17} investigated the chemical composition and multiplicity of 40 O4 - B0.5 stars with $v\,\sin{i} \geq 200$\,km\,s$^{-1}$ excluding SB2 binaries and Oe/Be stars. Out of 33 stars for which multi-epoch RV information could be derived, 5 were found to be SB1 binaries and another 9 were classified as RV-variables \citep{Caz17a}. Comparing the properties of these fast rotators with both single star and binary evolution models, \citet{Caz17} concluded that the properties of about 50\% of the stars could be explained by single star evolution. 
The properties of the other stars were likely affected by binary interaction. \citet{Caz17} pointed out that most stars appeared more He-enriched than expected from their N/O ratio. This property is hard to explain both with either single or binary evolution models.      

\citet{Mar17} investigated the chemical composition of 6 eclipsing SB2 O-star binaries. In 5 out of the 6 systems, the surface abundances were found to be only mildly affected by stellar evolution or mixing and did not differ from those of single stars. The majority of these systems were detached pre-interaction binaries, and \citet{Mar17} concluded that the effects of tides on the chemical evolution are limited. Moreover, the contact binary V382\,Cyg (O6.5\,V((f)) + O6\,V((f)), $P_{\rm orb} = 1.886$\,days) also did not display chemical enrichment. The only star showing a significant enrichment in their sample was the secondary of XZ\,Cep (O9.5\,V + B1\,III, $P_{\rm orb}$ = 5.097\,days). In this context, it is also interesting to note that the primary of XZ\,Cep rotates supersynchronously, whilst the opposite situation holds for the secondary.

\begin{figure}
\centering
\resizebox{16cm}{!}{\includegraphics{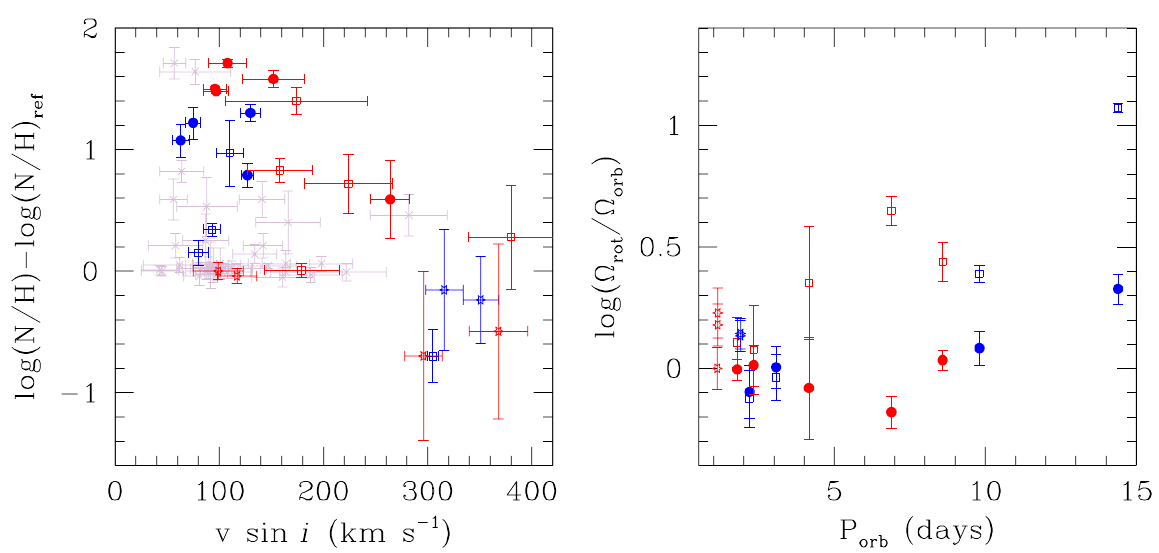}}
\bigskip
\begin{minipage}{12cm}
  \caption{Left: N/H abundance ratio with respect to the reference abundance (N/H = $0.60 \times 10^{-4}$ for Milky Way stars and N/H = $0.79 \times 10^{-5}$ for LMC objects) as a function of $v\,sin{i}$. The grey crosses stand for detached pre-interaction binary systems in the LMC \citep{Mah20a}. Blue and red symbols indicate interacting or post-interacting binaries respectively in our Galaxy and in the LMC. Filled dots stand for mass donors, whilst open squares represent accretors. Contact binaries are shown by stars. Right: ratio between rotational and orbital angular velocity for interacting or post-interacting binaries as a function of orbital period. The rotational axes are assumed to be perpendicular to the orbital plane. The symbols have the same meaning as in the left panel. Synchronisation due to tidal forces is clearly seen for $P_{\rm orb} \leq 4$\,days.\label{Hunter}}
\end{minipage} 
\end{figure}

\citet{Mah20a} investigated the atmospheric parameters of the components of 31 SB2 systems in the Tarantula region of the LMC. In agreement with \citet{Mar17}, they also concluded that tides play a minor role in the chemical enrichment of stars in detached pre-interaction binaries. Conversely, chemical enrichment was found in the spectra of mass donors of semi-detached systems, which were found to be mostly slow rotators. The mass gainers, some of which are rapid rotators spun up by transfer of angular momentum, showed only limited nitrogen enrichment. \citet{Mah20a} accordingly concluded that the outcome of binary interaction can explain the existence of slowly rotating N-rich O-stars as well as the existence of rapidly rotating non-enriched stars (see Fig.\,\ref{Hunter}). Both categories of objects are hard to understand in single star evolution scenarios.

\subsection{Magnetic fields}
As outlined in Sect.\,\ref{theory}, stellar mergers could lead to the formation of magnetic massive stars. According to this scenario, there should be no magnetic stars in close binary systems. This prediction agrees with the apparent dearth of magnetic stars in close binaries as found in dedicated surveys \citep{Nei15}. Still, there are a few short period binaries hosting magnetic stars. The most emblematic cases are the double magnetic system $\varepsilon$\,Lupi \citep[= HD\,136504, B3\,IV + B3:\,V, $P_{\rm orb} = 4.56$\,days, $e = 0.277$,][]{Shu15} and the rapidly rotating magnetic secondary star of Plaskett's Star \citep[][see also Sect.\,\ref{Plaskett}]{Lin08,Gru13}. The latter star has been interpreted as the mass gainer of a recent RLOF event. This opens up the possibility of a dynamo mechanism generated by differential rotation triggered by mass and angular momentum transfer \citep{Vanb18}. To observationally investigate the impact of past mass-transfer events on the generation of magnetic fields of massive stars, \citet{Naz17} analysed spectropolarimetry of a sample of 15 interacting or post-interaction massive binaries. No magnetic field was detected in any of them. This suggests that mass transfer events in massive binaries do not systematically trigger a stable strong magnetic field \citep{Naz17}. A radically different picture was drawn by \citet{Hub23}. These authors searched for Zeeman signatures in spectropolarimetry of 36 binary systems containing at least one O-type star. The systems in the sample had different evolutionary stages and covered a wide range of orbital separations. \citet{Hub23} reported apparent detections for 22 systems, thus suggesting that binarity would play a key role in the generation of magnetic fields. However, the systems detected by \citet{Hub23} include a number of binaries previously investigated by other groups who did not detect magnetic fields. The link between magnetic fields and past binary interactions is thus currently unclear and requires further studies. 

\citet{Fro24} considered the origin of the magnetic field of the Of?p star HD\,148937. This is a wide binary consisting of a mid- and a late-type O-star in a 26-year eccentric ($e \simeq 0.78$) orbit. A magnetic field was detected only for the hotter star. Despite the low amplitude of the RV variations, \citet{Fro24} reconstructed the spectra of the two stars via spectral disentangling. Both stars appear to be nitrogen-rich, although a quantitative assessment was only possible for the non-magnetic component. \citet{Fro24} compared the location of the stars in a Hertzsprung-Russell diagram to single star evolution models and found that the magnetic primary is apparently younger than the secondary star\footnote{The exact age difference depends on whether or not the N enrichment is taken into account.}. The primary star seems thus to be rejuvenated. \citet{Fro24} argue that this cannot be the result of a past RLOF episode because the donor star would then still be near filling its Roche lobe and the eccentricity should be low. They instead favour a merger scenario and argue that this could explain the extreme N abundance of the nebula which is more in line with a WN composition than with the photospheric abundances of the stars \citep{Mah17,Lim24}. According to this scenario, HD\,148937 was originally a triple system with a close inner binary that underwent a merger event during which the magnetic field was produced and the nebula was ejected.

\subsection{Overluminosities}
\begin{figure}[htb]
\centering
\resizebox{16cm}{!}{\includegraphics{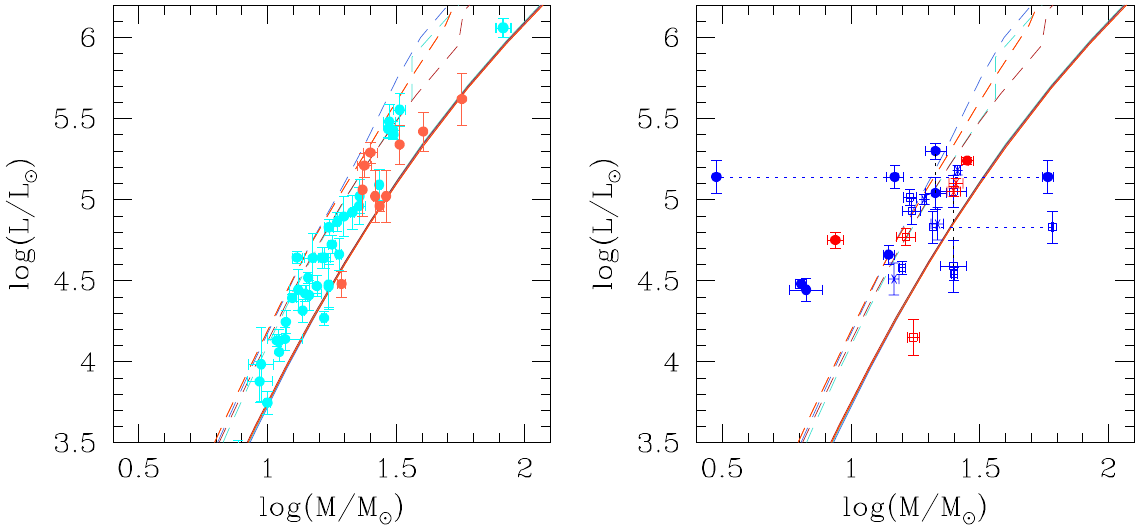}}
\bigskip
\begin{minipage}{12cm}
  \caption{Left: mass-luminosity relation for the components of detached pre-interaction eclipsing massive binaries. The cyan and red symbols correspond to binary systems in our Galaxy and in the LMC. Theoretical mass-luminosity relations are illustrated at the onset (solid lines) and end (dashed lines) of the core hydrogen burning phase. The different lines correspond to models with metallicity Z = 0.014 without rotation (turquoise) and with initial rotation of 40\% critical (blue), as well as with Z = 0.006 without rotation (dark red) and 40\% critical rotation (orange). Right: mass-luminosity relation for the components of contact binaries (stars) as well as for the mass donors (filled circles) and accretors (open boxes) of semi-detached or post-RLOF systems. Blue (resp.\ red) symbols indicate binaries in our Galaxy (resp.\ the LMC). The same theoretical relations are shown by the lines. The dotted horizontal lines connect the possible locations of the components of Plaskett's Star (see Sect.\,\ref{Plaskett}). The dotted vertical lines indicate the possible solutions for the 29\,CMa system. \label{ML}}
\end{minipage} 
\end{figure}
Figure\,\ref{ML} compares the mass-luminosity relation of the components of interacting or post-interaction binaries (right panel) with that of core-hydrogen burning massive stars in detached eclipsing binaries (left panel). To compare with theoretical models of single stars, we use the mass-luminosity relation of single rotating main-sequence stars at the onset and at the end of the core-hydrogen burning phase. These relations are taken from the Geneva models for solar and LMC metallicities \citep{Eks12,Egg21}. Regarding observational data, we assembled a sample of detached, pre-interaction eclipsing massive binaries in our Galaxy and in the LMC. The main bibliographic sources are the compilation of \citet[][and references therein]{Eker14} complemented by several additional studies \citep{Fre01,Rauw01,Mas02,Rauw05,Mah20b,Ros20a,Ros22a,Ros22b}. The left panel of Fig.\,\ref{ML} shows that the components of detached, unevolved massive binaries nicely follow the theoretical relation. The situation is more complicated when we consider evolved systems (right panel). Here we focus on systems with relatively well determined parameters: LZ\,Cep \citep{Mah11b}; V382\,Cyg, V448\,Cyg, and XZ\,Cep \citep{Har97}; VFTS\,061, VFTS\,176 and VFTS\,352 \citep{Mah20b}; and TU\,Mus \citep{Lin07}. We further include two systems with larger uncertainties: Plaskett's Star (see Sect.\,\ref{Plaskett}) and 29\,CMa (see Sect.\,\ref{How}). On average, the mass donors appear overluminous compared to main-sequence stars of same mass. However, whilst this overluminosity can reach values of 1\,dex or more, not all mass donors display an overluminosity. The mass gainers display large dispersions in luminosity, with some of them being underluminous. Finally, the components of contact systems in Fig.\,\ref{ML} do not exhibit systematic over- or underluminosities. This latter result is at odds with the conclusion of \citet{Abd21} who found systematic overluminosities in three contact binaries. These discrepant conclusions probably reflect the huge uncertainties and biases that affect our knowledge of the properties of these objects (see Sect.\,\ref{How}). 

\section{How well do we know evolved massive binaries? \label{How}}
From the examples presented in Sect.\,\ref{population}, we conclude that whilst binary interactions impact the properties of massive stars populations, none of the observational diagnostics taken individually can be unambiguously attributed to the sole effect of previous binary interaction phases. In this section, we thus review a handful of massive binaries that are considered good examples of post-RLOF systems, and we critically ask ourselves how well we know and understand these systems.

Modern observational studies of massive binaries usually rely on different types of data (photometry, spectroscopy, (spectro)polarimetry, interferometry, etc.). Spectral disentangling is applied to phase-resolved spectroscopy to reconstruct the individual spectra of the components and to infer their epoch-dependent RVs. Whilst the RVs are used to build the orbital solution of the system, the reconstructed spectra are analysed with non-LTE stellar model atmosphere codes to infer the stellar properties and the surface chemical compositions of the stars. In parallel, high-precision space-borne photometry, collected with missions such as {\it CoRoT}, {\it Kepler}, {\it BRITE} or {\it TESS}, is adjusted with sophisticated binary light curve models. The stellar parameters obtained this way can then be compared to the predictions of theoretical models. 
\subsection{Contact binaries}
\begin{figure}
\centering
\resizebox{12cm}{!}{\includegraphics{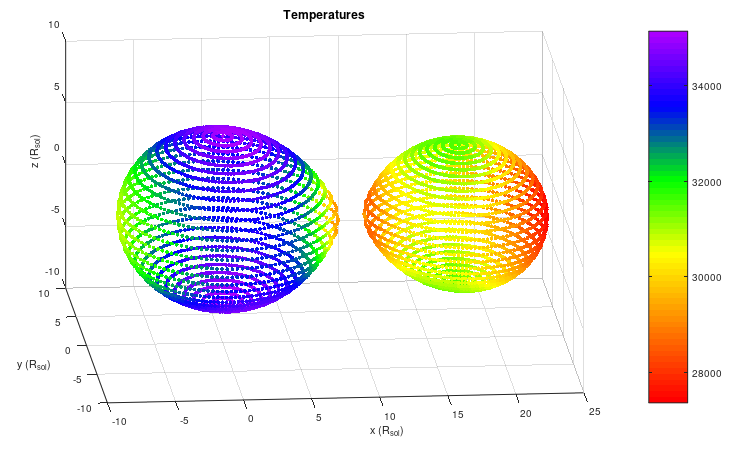}}
\bigskip
\begin{minipage}{12cm}
  \caption{Simulated surface temperature distribution of the TU\,Mus system computed with the CoMBiSpeC code \citep{Pal12,Pal13}. The calculations account for the effect of mutual heating and radiation pressure. \label{TUMus}}
\end{minipage} 
\end{figure}

TU\,Mus (HD\,100213) is an O-star eclipsing binary with a period of 1.387\,d. Using optical spectra, \citet{Lin07} classified the system as O7.5\,V + O9.5\,V, whereas \citet{Pen08} assigned earlier O7\,V + O8\,V spectral types based on {\it IUE} UV spectra. Analyses of the {\it Hipparcos} light curve of this system indicate a contact configuration \citep{Lin07,Pen08}. The current mass ratio is about $M_p/M_s = 1.45$ according to \citet{Lin07} or 1.60 according to \citet{Pen08}. The system is thought to experience a slow Case A RLOF where the mass gainer (the currently more massive star) also fills its Roche lobe due to its increase in mass and the rejuvenation of its interior \citep{Pen08}. Both stars appear to be rotating nearly synchronously with the orbital motion \citep{Pen08}, as expected for a situation where the orbital period is short and the tidal forces are strong (see right panel of Fig.\,\ref{Hunter}).
A remarkable feature of TU\,Mus is that different optical spectral lines yield different amplitudes of the RV curves \citep{Lin07} and that UV spectra yield RV amplitudes ($K_p$ and $K_s$) that are significantly lower than those obtained from optical spectra \citep{Pen08}. For instance, the He\,{\sc i} $\lambda$\,5876 line displays a $K_p$ value that is 30\,km\,s$^{-1}$ (i.e.\ 12\%) larger than the corresponding value of the He\,{\sc ii} $\lambda$\,5412 line. A very similar 10\% discrepancy holds for the RV amplitude of the secondary. Compared with UV spectra, the differences are even higher. This situation reflects the highly non-uniform surface temperature distributions due to the impact of gravity darkening of the highly distorted stars in this contact binary and mutual heating effects \citep[][see also Fig.\,\ref{TUMus}]{Pal13}. The result is a shift of the centres of light of different lines with respect to the centre of gravity of the stars. Such a situation has a severe negative impact on our knowledge of the masses of the stars. Indeed, the masses (16.7 and 10.4\,M$_{\odot}$) inferred by \citet{Pen08} are about 30\% lower than the values (21.7 and 14.7\,M$_{\odot}$) found by \citet{Lin07} from the mean of their optical RV curves. This is a huge and unfortunate uncertainty when it comes to comparison with theoretical models. Moreover, the fact that different lines follow different RV curves hampers our possibility to apply spectral disentangling to this system, and thereby prevents us from studying the chemical compositions of the stars. 

29\,CMa (= HD\,57060, $P_{\rm orb} = 4.39$\,days) is another evolved O-star binary system which is thought to be in a contact or semi-detached configuration \citep{Bag94,Ant11}. The primary is an O8.5\,I supergiant with a relatively powerful stellar wind as indicated by the rather strong N\,{\sc iii} $\lambda\lambda$\,4634-40 and He\,{\sc ii} $\lambda$\,4686 emission lines. The secondary spectral signature is very weak. Based on optical spectra, \citet{Linder} derived an O9.7\,V spectral type for the secondary, a mass ratio of $M_s/M_p = 1.16$, and an optical brightness ratio $l_s/l_p = 0.21 \pm 0.08$. This latter value is at odds with the contact configuration inferred from the {\it Hipparcos} light curve of the system \citep{Linder,Ant11}, which rather yields $l_s/l_p = 1.08$ for a mass ratio of 1.16. Whilst the disentangled spectra of \citet{Linder} clearly point at a nitrogen enrichment of the primary star, the huge uncertainty about the brightness ratio prevented a quantitative assessment of this effect. This is because the brightness ratio is a key parameter for assessing the dilution correction of the line strength in the normalized disentangled spectra. Similar issues have been encountered for several other evolved massive binaries. Solving this problem requires a revision of the spectroscopic and photometric analyses. The effects of the primary stellar wind might play a key role here, and could eventually change our interpretation of the photometric light curve. The existence of a strong stellar wind and the nitrogen enrichment suggest that the primary is about to become a WR star.

\citet{Abd21} investigated three overcontact binaries: V382\,Cyg (= HD\,228854, $P_{\rm orb} = 1.886$\,days) in the Milky Way, VFTS\,352 ($P_{\rm orb} = 1.124$\,days) in the LMC, and OGLE SMC-SC10\,108086 ($P_{\rm orb} = 0.883$\,days) in the Small Magellanic Cloud (SMC). They found that both components of these systems are highly overluminous, but they observed no evidence for altered He and CNO surface abundances \citep[confirming a result previously found for V382\,Cyg by][]{Mar17}. Moreover, in unequal-mass systems, they noted that the temperature of the less massive component is shifted towards that of the more massive star, and exceeds the temperature expected for the mass of the star. \citet{Abd21} proposed that this could result from irradiation by the primary, efficient heat exchanges through the common envelope, or efficient internal mixing. Only the latter mechanism can account for the global overluminosity (i.e., of both the primary and the secondary stars, see however Fig.\,\ref{ML} which does not reveal a systematic overluminosity). However, it is at odds with the lack of chemical enrichment. \citet{Abd21} argue that this could be an indication of a non-conservative mass loss. While this could explain why the mass gainer does not show enrichment (lack of pollution by the material of the mass donor), this explanation appears less convincing regarding the mass donor whose envelope would be stripped off. For V382\,Cyg and OGLE\,SMC-SC10\,108086, \citet{Abd21} found that spectral lines are broader than expected for stars which would be in synchronous rotation. However, no such effect was reported by \citet{Mar17} for V382\,Cyg. 

Another interesting system in this context is LSS\,3074 (= V889\,Cen). The components of this binary are classified as O4.5\,If$^+$ + O6.5-7\,If, and the orbital period is $P_{\rm orb} = 2.185$\,days \citep[][and referencese therein]{Raucq17}. The photometric light curve displays ellipsoidal variations with a depth of 0.2\,mag, but lacks clear photometric eclipses. Conventional binary light curve models can only fit this light curve assuming a contact or overcontact configuration \citep{Raucq17}. The best-fit orbital inclination ($i = 54.5^{\circ}$) then implies surprisingly low masses for the two stars ($M_p = (14.8 \pm 1.1)$\,M$_{\odot}$ and $M_s = (17.2 \pm 1.4)$\,M$_{\odot}$). Both stars would thus be hotter and more luminous than expected for their masses. Both stars display a strong N enrichment in their spectrum. The O4.5 primary also displays an enhanced He surface abundance. The analysis of the light curve implies an optical brightness ratio of $l_1/l_2 = 1.09$. As for 29\,CMa, this value is clearly at odds with the spectroscopic brightness ratio of $2.50 \pm 0.43$. This latter value indeed requires a semi-detached or detached configuration. Comparing the properties of LSS\,3074 with binary evolutionary tracks, \citet{Raucq17} suggested that the system is probably undergoing a highly non-conservative slow Case B mass transfer.  

\citet{Ric23} analysed the ON3\,If$^*$ + O5.5\,V((f)) binary system NGC\,346 SSN\,7 ($P_{\rm orb}$ = 3.07) in the SMC. From the orbital solution, they showed that the secondary star is the more massive object ($M_p/M_s = 0.68$). The dynamical minimum masses of the stars are low, suggesting a low inclination. Using a model atmosphere code, \citet{Ric23} inferred spectroscopic masses of 32 and 55\,M$_{\odot}$ for the primary and secondary, respectively. This analysis further revealed that the primary's atmosphere is N-rich and depleted in O, as expected from the ON spectral classification. The chemical compositions and the mass ratio below 1.0 suggest a past mass-exchange episode. \citet{Ric23} conclude that the system is in a contact configuration undergoing a slow Case A mass exchange.

\subsection{Semi-detached binaries}
\begin{figure}
\centering
\resizebox{12cm}{!}{\includegraphics{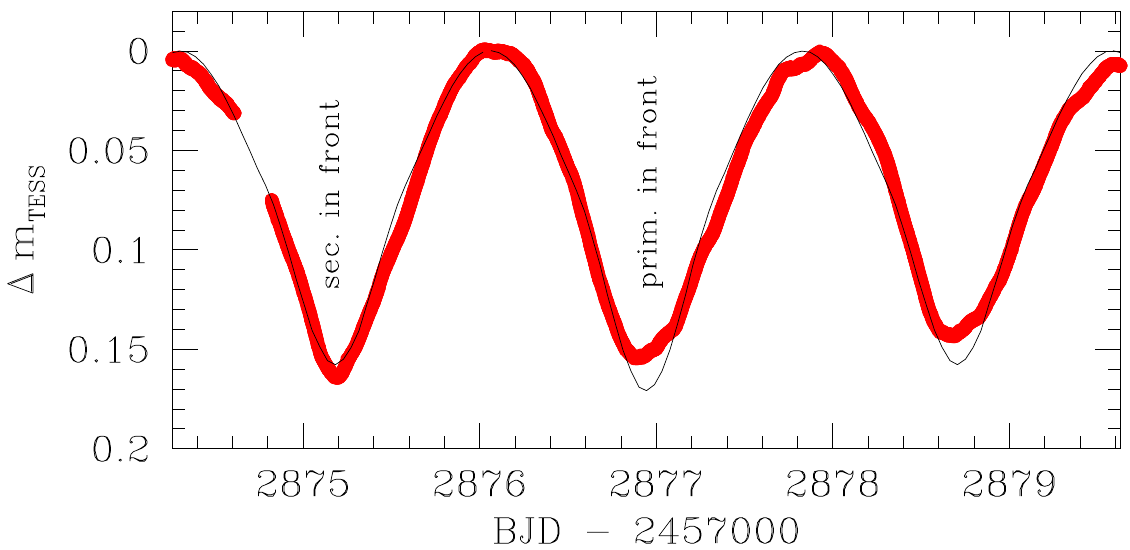}}
\bigskip
\begin{minipage}{12cm}
 \caption{Light curve of AO\,Cas from {\it TESS} sector 57. Observed magnitudes are shown by red dots, whilst the black solid line yields a preliminary best-fit curve adjusted to the subset of observations taken between BJD\,2459874.26 and 2459879.6. The fitted light curve leads to an inclination of $i = 55.7^{\circ}$. The primary fills its Roche lobe, whilst the secondary has a filling factor of 0.97.\label{AOCas}}
\end{minipage} 
\end{figure}

AO\,Cas (= HD\,1337) has an orbital period of 3.52\,days. Based on disentangling of optical spectra, \citet{Linder} derived a spectral classification O9.5\,III for the primary and O9\,V for the secondary. The secondary is currently 1.6 times more massive than the primary star, but would be about four times fainter in the optical than the primary. \citet{Linder} analysed {\it Hipparcos} photometry of AO\,Cas and found a solution with an inclination $i = 65.7^{\circ}$. This solution yields rather low masses for both stars ($M_p = 9.6$ and $M_s = 15.6$\,M$_{\odot}$). The photometric brightness ratio of the {\it Hipparcos} solution of \citet{Linder} agrees with the spectroscopic value, and the best-fit model of the light curve consists of ellipsoidal variations and grazing eclipses. However, the observations display substantial scatter around the model. 
The {\it TESS} light curve of AO\,Cas seems to tell a different story though (see Fig.\,\ref{AOCas}). It is dominated by ellipsoidal variations with no indication of grazing eclipses. The light curve further exhibits additional variations with amplitudes up to 0.02\,mag. These variations affect the determination of the stellar parameters. Nonetheless, a preliminary analysis of the light curve indicates that the ellipsoidal variations are best explained if the more massive secondary also fills a significant fraction of its Roche lobe. Whilst the lower inclination obtained from the {\it TESS} data leads to larger masses (12.9 and 20.9\,M$_{\odot}$), it now implies a tension between the spectroscopic optical brightness ratio and the photometric value. Indeed, the new photometric value is equal to $l_s/l_p = 1.62$, which differs significantly from the spectroscopic value \citep[$l_s/l_p = 0.26 \pm 0.16$,][]{Linder}.   

LZ\,Cep (= HD\,209481) is an O9\,III + ON9.7\,V binary with an orbital period of 3.07\,days \citep[][and references therein]{Mah11b}. The ON spectral type of the secondary hints at a strong N enrichment which was confirmed, along with a strong He enrichment and a C depletion, by the quantitative analysis of \citet{Mah11b}. LZ\,Cep is in a semi-detached configuration as indicated by the analysis of the ellipsoidal variations seen in the {\it Hipparcos} light curve, and it is most likely the secondary star that fills-up its Roche lobe. The O9\,III primary star has a mass of 16\,M$_{\odot}$, whilst the ON secondary has a mass of only 6.5\,M$_{\odot}$. \citet{Mah11b} therefore suggest that the secondary is a core He-burning object which has retained some parts of its envelope and has a comparatively weak wind. Given the lack of strong chemical enrichment of the primary star, these authors conclude that the mass transfer must have been quite inefficient. In view of the orbital period, tidal interactions are expected to be strong, and both stars are indeed found to have nearly synchronized rotational and orbital periods (see Fig.\,\ref{Hunter}).

\subsection{Detached systems}
Some detached massive binary systems display clear hints of past RLOF episodes. \citet{Raucq16} analysed the disentangled spectra of the O7.5\,If + ON9.7\,I system HD\,149404 ($P_{\rm orb} = 9.81$\,days, $M_s/M_p = 0.63$). As suggested by the ON spectral type of the secondary, this star presents a strong nitrogen enrichment and carbon depletion. Its [N/C] ratio was found to be 150 times solar, whilst that of the (currently) more massive primary was only 5 times solar. This hints at a system which has undergone a past Case A RLOF episode although {\it BRITE} photometry suggests that the secondary star is still close to filling its Roche lobe \citep{Rauw19}. There are some caveats though. Indeed, the N enrichment of the donor star should go along with an increase of the He abundance, which was not observed \citep{Raucq16}. Moreover, the observed enrichment would be more typical of Case B mass transfer whilst the properties of the system suggest a Case A event. Last but not least, the uncertainties on the orbital inclination \citep[21 -- 31$^{\circ}$,][]{Rauw19} imply uncomfortably large uncertainties on the masses. Nonetheless, at least the secondary star appears to be overluminous for its mass.

Another candidate detached post-RLOF system is the eclipsing binary AzV\,476 in the SMC \citep[O4\,IV-III((f))p + O9.5:Vn, $P_{\rm orb} = 9.37$\,days, $e = 0.24$,][]{Pau22}. The spectrum of the O4\,IV-III primary star displays a significant N enrichment. Moreover, with a mass of only about 20\,M$_{\odot}$, it appears clearly too hot for its mass. The secondary has a mass of 18\,M$_{\odot}$ which is well in line with its spectral type, but displays a very high projected rotational velocity \citep[$v_s\,\sin{i} = 425$\,km\,s$^{-1}$,][]{Pau22}. Comparing the properties of this system with binary evolutionary models, \citet{Pau22} concluded that AzV\,476 must have experienced a Case B RLOF in the past, and that the primary is likely a core He burning object which has lost about half of its initial mass. 

Finally, another interesting example of a post-RLOF system could be the WN6o + O9.5-9.7\,V-III binary WR\,138 \citep[= HD\,193077, $P_{\rm orb} = 1559$\,days, $e = 0.16$,][]{Rauw23}. The mass ratio $M_{\rm WN6o}/M_{\rm OB} = 0.53 \pm 0.09$ and the highly asynchronous rotation of the OB star ($v_{\rm OB}\,\sin{i} = (350 \pm 10)$\,km\,s$^{-1}$ suggest that this system evolved through a Case B mass and angular momentum exchange episode in a way similar to the scenario of \citet{Vanb18}. WR\,138 has a significantly longer orbital period than the systems studied by \citet{Sha17,Sha20} which had periods ranging between 1.9 and 78.5\,days. Hence, WR\,138 provides a more extreme example of a binary system where tidal interactions are unable to explain the lower than critical rotation rate of the OB star.

\subsection{Revisiting Plaskett's Star\label{Plaskett}}
HD\,47129 (aka Plaskett's Star) has been considered as a kind of textbook example of the outcome of Case A mass transfer, apparently ticking nearly all the boxes regarding evidence of a past RLOF episode \citep[][and references therein]{Lin08}. This system consists of an O8\,III/I star with narrow spectral lines (hereafter the primary star) and an O7.5\,III companion (the secondary) displaying significantly broader lines. The primary lines describe a well documented circular orbital motion with a period of 14.396\,days and a mass function of 12.3\,M$_{\odot}$. Due to its rapid rotation and line profile variability, the RVs of the secondary star are much more difficult to measure. \citet{Lin08} used a spectral disentangling method to measure the RVs of both components and derived a mass ratio of $\frac{M_s}{M_p} = 1.05$, which results in both stars having minimum current masses around 45\,M$_{\odot}$. The lines of the primary component clearly exhibit the signature of a N overabundance \citep[by a factor $16.6 \pm 5.0$ with respect to solar,][]{Lin08}. These properties were interpreted as the outcome of a recent Case A \citep[or Case B, according to][]{Vanb18} mass and angular momentum transfer with the primary being the partially stripped mass donor and the secondary being the mass gainer that was spun-up by accretion.

Analysing the {\it CoRoT} light curve of HD\,47129, \citet{Mah11a} discovered photometric variations related to the orbital cycle as well as a modulation with a 0.82\,d$^{-1}$ frequency along with six harmonic frequencies. The orbital light curve was interpreted as ellipsoidal variations altered by a bright spot on the primary facing the secondary and resulting from the wind-wind collision. The 0.82\,d$^{-1}$ frequency, which was subsequently also detected in spectroscopy \citep{Pal14}, and its harmonics were tentatively associated with non-radial pulsations \citep{Mah11a}. Using high-resolution spectropolarimetry, \citet{Gru13} subsequently discovered a strong (2.8\,kG) magnetic field associated with the secondary star. These authors suggested that the 0.82\,d$^{-1}$ frequency would be the rotation frequency of the secondary star.  

\citet{Gru22} confirmed the connection of the 0.82\,d$^{-1}$ frequency to the strength variations of the H$\alpha$, H$\beta$, H$\gamma$ and He\,{\sc ii} $\lambda$\,4686 emission components which were shown to arise in the magnetosphere of the secondary star. Therefore, this frequency corresponds indeed to the rotational frequency of the magnetic secondary star. This star is thus the first example of an O-star hosting a rigidly rotating centrifugal magnetosphere. Most importantly, \citet{Gru22} used Zeeman Doppler Imaging to map the magnetosphere of the secondary star of HD\,47129. The results of such a reconstruction depend on the secondary's RVs. \citet{Gru22} found that adopting the secondary RV amplitude from \citet{Lin08} led to a rather poor adjustment of the Stokes $V$ spectra and a complex magnetic field morphology. Instead, the formally best-fitting RV amplitude of the secondary found in the Zeeman Doppler Imaging was about 30\,km\,s$^{-1}$, roughly one sixth of the value found by \citet{Lin08}.

If confirmed, this reduced RV amplitude leads to a major shift in our interpretation of this system. Let us briefly consider the implications of this result. Adopting the 30\,km\,s$^{-1}$ RV amplitude for the secondary, as found by \citet{Gru22}, implies a mass ratio $\frac{M_s}{M_p} = 6.7$. Together with the primary's mass function, established by \citet{Lin08}, we obtain $M_s\,\sin^3{i} = (16.2 \pm 0.7)$\,M$_{\odot}$. For the formally best fit orbital inclination of $67^{\circ}$ obtained by \citet{Mah11a}, the absolute masses of the components become $M_p \sim 3$\,M$_{\odot}$ and $M_s = 20.8$\,M$_{\odot}$, and the orbital separation amounts to $a = 71.8$\,R$_{\odot}$. If we adopt the distance ($1283^{+127}_{-118}$\,pc) from the third {\it GAIA} data release \citep{Bai21} and the spectroscopic optical brightness ratio of $l_p/l_s = 1.9 \pm 0.1$, we can estimate absolute magnitudes of $M_{{\rm V}, p} = -4.98 \pm 0.23$ and $M_{{\rm V}, s} = -4.28 \pm 0.23$. With the bolometric corrections from \citet{Mar06} and the effective temperatures inferred by \citet{Lin08}, we then derive radii of $R_p = 11.0$\,R$_{\odot}$ and $R_s = 8.0$\,R$_{\odot}$. With the projected rotational velocities \citep[$v_p\,\sin{i} \sim 75$\,km\,s$^{-1}$ and $v_s\,\sin{i} \sim 305$\,km\,s$^{-1}$,][]{Lin08} and assuming $i = 67^{\circ}$ \citep{Mah11a}, the rotational periods of the primary and secondary star become 6.80 and 1.22\,days, respectively. The latter value perfectly matches the 0.82\,d$^{-1}$ frequency.

Assuming the mass ratio estimated by \citet{Gru22} is confirmed by future studies, the revised picture of Plaskett's Star would be the following: the primary star was the initially more massive star and has transferred considerable amounts of mass and angular momentum to its companion during a Case A RLOF phase. The companion was spun up significantly during this process. The reversal of the mass ratio led to an expansion of the orbit, thereby ending the mass transfer episode. Both stars are now well within their Roche lobes. The stripped primary star is now the object that we need to understand! Indeed, this star would be highly overluminous for its mass. Such overluminosities are expected for stripped massive stars, but the primary star of HD\,47129 does not yet look like a genuine Wolf-Rayet star. However, with a luminosity of $\log{\frac{L_{{\rm bol},\,p}}{L_{\odot}}} = 5.13 \pm 0.09$, an effective temperature near 33500\,K, and a mass of $\sim 3$\,M$_{\odot}$, its properties (especially its mass over radius ratio) strongly differ from those of normal O-type stars. Moreover, given its radius, one could expect the primary to appear as a bloated stripped star, which is not the case either. Indeed, beside the observed temperature which would be quite high for such an object, one important discrepancy concerns the value of $\log{g}$ which would be $\sim 2.8$ with the above mass and radius, whilst the model atmosphere fit presented by \citet{Lin08} rather led to $3.5 \pm 0.1$. Another puzzle to solve would be to find out how the primary star acquired its apparently super-synchronous rotation. 

\section{Conclusions}
Various observational properties of massive binaries are best explained by ongoing or past binary interaction phases. The most prominent smoking guns are rapid rotation, enhanced nitrogen and helium abundances, and luminosities that do not match the stellar mass. However, taken individually, at least some of these properties might also be the outcome of single star evolution. Hence, detecting only one of these properties is not sufficient to draw conclusions about past mass exchange episodes. At first glance, observational investigations of specific binaries are in broad agreement with the predictions of binary evolution models. However, the devil often lies in the details! In-depth multi-technique investigations of such systems frequently lead to surprising or even conflicting results, implying significant uncertainties in our knowledge of their fundamental properties. Whilst some of these problems hint at some missing physics in the binary evolution models, the majority probably reflect deficiencies in the current analyses of observational data. In some systems, even such fundamental parameters as the stellar masses are subject to uncomfortably large uncertainties. It is thus highly desirable to continue investigating those systems in the coming years. In some cases, existing analysis tools might not be sufficient to achieve a full understanding and more sophisticated approaches will probably have to be developed to account for the specificities of these systems.  


\begin{furtherinformation}
\begin{orcids}
  \orcid{0000-0003-4715-9871}{Gregor}{Rauw}
\end{orcids}


\begin{conflictsofinterest}
The author declares no conflict of interest.
\end{conflictsofinterest}

\end{furtherinformation}

\bibliographystyle{bullsrsl-en}

\bibliography{G_Rauw_2_deb}

\end{document}